\documentclass{amsart}
\usepackage[margin=2.7cm]{geometry}
\usepackage{amsfonts}
\usepackage{amsmath}
\usepackage{mathtools}
\usepackage{braket}
\usepackage{amsthm}
\usepackage{xcolor}
\usepackage{rotating}
\usepackage[numbers]{natbib}
\usepackage{float}
\usepackage{url}
\usepackage[english]{babel}

\usepackage[pdfusetitle]{hyperref}

\begin{document}

\title{The Quantum Mechanical Problem of a Particle on a Ring with Delta Well}

\author{Raphael J.F.~Berger}
\address{Division of Chemistry and Physics of Materials, Paris-Lodron University Salzburg, Salzburg, A-5020, Austria}
\email{raphael.berger@sbg.ac.at}
\thanks{RJFB acknowledges funding from DFG (German Research Foundation) within the Priority Program SPP1807 
“Control of LD in molecular chemistry”, grant BE4632/2-2, project no. 271386299)}

\begin{abstract}

  The problem of a spin-free electron with mass $m$, charge $e$  confined onto a
  ring of radius $R_0$ and with an attractive Dirac
  delta potentential with scaling factor (depth) $\kappa$
  in non-relativistic theory has closed form analytical solutions.
  The single bound state function is of the form of a hyperbolic
  cosine that however contains a paramter $d>0$  which
  is the single positive real solution of the transcentdental
  equation $\coth(d) = \lambda d$ for non
  zero real $\lambda=\frac{2}{\pi\kappa}$. The energy eigenvalue of the bound state $\varepsilon=-\frac{d^2}{2\pi^2}\approx \frac{q e m R_0}{2 \hbar^2}$. In addition a discrete inifinty of
  unbounded solutions exists, formally they are obtained from the terms for the bound solution by
  substituting $d \to i d $ yielding $\cot(d) = \lambda d$ as characteristic equation with the
  correspondig set of solutions $d_k, k\in\mathbb{N}$, the respective state functions obtain via
  $\cosh(x)\overset{x \to i x}{\longrightarrow}\cos(x)$ of the form of cosine functions.
\end{abstract}

\date{\today}

\maketitle
\section{Introduction}

There is only roughly a dozen of quantum mechanical (QM) systems with an
analytical solution.\cite{enwiki:1121884573} QM problems with analytical solutions are not only
of great didactical use to demonstrate how such problems can be solved but o
ften serve also as physical toy model systems for otherwise unsovable problems which however share
the principal characteristics.

One example is the particle in the Dirac delta potential\cite{wiki_dirac} which not ony
serves with its bound state as an one-dimensional analog of the hydrogen atom but can also be
interpreted as the simplest model for electron scattering when one regards the unbound
states and allows for simple calculation of reflection and transmissoin rates on step- and
related potentials, which is for example of relevance for the theory of scanning tunnel microscopy.

In the context of our research on symmetry breaking in rotationally invariant
systems\cite{Antiaro} we came across the analogous problem where the particle is but
confined to an atomic scale ring. As it turned out that the solution of this QM model system
has not yet been described in detail the literature (implicitly this system is contained in a
work on a closely related Berry-phase model\cite{Berry}), we report on our 
results in the following.

\section{Solution}
\subsection{Schr\"odinger equation}
A spin-free electron ({\em i.e.} a particle with charge $-e$ and mass $m_e$) 
on in a ring shaped space with radius $R_0$ and a $\delta$ function well of amplitude
$-q e$ corresponding to an attractive potential with an integrated total
charge of $+q e$ 
is regarded in non-relativistic quantum mechanic theory. \\

As is well known from textbooks the
Hamiltonian for the particle on a ring of radius $R_0$ in 2D polar 
coordinates ($r\in\mathbb{R}^+,\vartheta\in[0,2\pi)$) in SI units is given by 
\begin{align}
\hat{H'} & = -\frac{\hbar^2}{2 m R_0^2} \frac{\partial^2}{\partial \vartheta^2}
\label{h-ring}
\end{align}
The ring shall contain an {\em attractive} potential $\hat{V}$ with respect to the electron
in the form of a Dirac-$\delta$ function\footnote{The dimension of the argument of the $\delta$
function has to be coosen such as $\int \delta \;dx = 1$ is
dimensionless} and the potential shall integrate over the
whole space ($R_0\times[0,2\pi)$) to the product of electron and potential charge of $-e q<0$:  
\begin{align}
\hat{V} & = -q e \delta((\vartheta-\vartheta_0)l), 
\label{pot}
\end{align}
were $\vartheta$ is given in units of $\mathrm[rad]$, thus
formally $l = 1 \mathrm{rad}^{-1} = 1 \frac{1m}{1m} = 1$
and hence can be dropped in the following. 
Without loss of generality we will set $\vartheta_0 = 0$ such that the
Schr\"odinger equation for the problem becomes
\begin{align}
\hat{H}\psi(\vartheta) &  = E \psi(\vartheta) \nonumber \\ 
(\hat{H'}+\hat{V}) \psi(\vartheta)   &  = E \psi(\vartheta) \nonumber \\
 \left( -\frac{\hbar^2}{2 m R_0^2}\frac{\partial^2}{\partial \vartheta^2} -
 q e \delta(\vartheta)\right)\psi(\vartheta) & = E \psi(\vartheta) \label{SE-1} \\
% \left( -\frac{\hbar^2}{2 m R_0^2}\frac{\partial^2}{\partial \vartheta^2} -
% q e \delta(\vartheta)\right)\psi(\vartheta) - E \psi(\vartheta) & = 0 \\
% \left( \frac{\partial^2}{\partial \vartheta^2} -
% q e (-\frac{\hbar^2}{2 m R_0^2})^{-1} \delta(\vartheta)\right)\psi(\vartheta) - E (-\frac{\hbar^2}{2 m R_0^2})^{-1} \psi(\vartheta) & = 0\\
% \left( \frac{\partial^2}{\partial \vartheta^2} +
% q e (\frac{2 m R_0^2}{\hbar^2}) \delta(\vartheta)\right)\psi(\vartheta) + E (\frac{2 m R_0^2}{\hbar^2}) \psi(\vartheta) & = 0 \\
% \left( \frac{\partial^2}{\partial \vartheta^2} +
% q e (\frac{2 m R_0^2}{\hbar^2}) \delta(\vartheta) + E (\frac{2 m R_0^2}{\hbar^2})\right)\psi(\vartheta)  & = 0 \\
 \left( \frac{\partial^2}{\partial \vartheta^2} +
 \frac{2 q e m R_0^2}{\hbar^2} \delta(\vartheta) + E \frac{2 m R_0^2}{\hbar^2}\right)\psi(\vartheta)  & = 0 \nonumber
\end{align}
For simplicity we combine the constants in
\begin{align}
\kappa =  \frac{2 q e m R_0^2}{\hbar^2} \label{kappa}
\end{align}
and
\begin{align}
  \epsilon =  \frac{2 m R_0^2}{\hbar^2} E \label{epsilon}
\end{align}
such that we obtain
\begin{align}
 \psi'' + (\epsilon + \kappa\delta)\psi & = 0 
\label{SE}
\end{align}
\subsection{Bound state ($E<0$)}
The assumption of $(E-V)<0$ or (ignoring the singularity at the
origin) $E<0$, or $\epsilon<0$, respectively thus leads to the bound state solutions.
For the symmetry of the problem and the form of the differential
equation (\ref{SE}) we chose the {\em Ansatz}
\begin{align}
\psi(\vartheta;d) = N' (e^{-d\vartheta} + e^{d(\vartheta - 2\pi)}) = N \cosh(d(x-\pi))
\label{ansatz}
\end{align}
Yielding the normalization constant
\begin{align}
%  C' = \left(\frac{e^{4 \pi  d}-1}{d} + 4 \pi  e^{2 \pi  d}\right)^{-\frac{1}{2}}
  N = \sqrt{\frac{\sinh (2 \pi  d)}{2 d}+\pi}^{-1}
\label{norma}
\end{align}

To determine the exponent $d$ one in principle has to insert
(\ref{ansatz}) into (\ref{SE}) and attempt to match $d$ to the
boundary conditions. One boundary condition, the symmtry of
the system, was already accounted for choosing the same exponent $d$
for both functions but with different sign. Since a $\delta$-function
is appearing one has to chose an appropriate strategy to evaluate the
result of combining (\ref{ansatz}) and (\ref{SE}). The strategy is to 
integrate the Schr\"odinger equation in an $\epsilon$ ball around the
origin of the $\delta$-function and to perform the limit of
$\epsilon\to 0$, this results in
\begin{align}
\lim_{\epsilon\to 0}\int_{0-\epsilon}^{0+\epsilon}  \left(\frac{\partial^2}{\partial \vartheta^2} \psi(\vartheta) \right) +
 \kappa\delta(\vartheta)\psi(\vartheta) - E' \psi(\vartheta) \; d \vartheta & =
 \lim_{\epsilon\to 0}\int_{0-\epsilon}^{0+\epsilon} 0 d \vartheta \nonumber \\
\psi'(0^+) - \psi'(0^-) + \kappa\psi(0) & = 0
\label{SE_integral}
\end{align} 
Inserting (\ref{ansatz}) in (\ref{SE_integral}) yields
\begin{align}
-2 d + 2 d e^{-2\pi d} + \kappa (1+e^{-2\pi d}) & = 0 \nonumber \\
\coth(\pi d) & = \frac{2}{\kappa}d
\label{bcond_1}
\end{align} 
using $d'=\pi d$ and $\lambda=\frac{2}{\pi \kappa}$ we obtain
\begin{align}
\coth(d') & = \lambda d'
\label{bcond}
\end{align} 
here for $\lambda > 0$ and real $d'$ exactly one pair of solutions $d'_+ = - d'_-$  exists and yielding the same
function $\psi(\vartheta;d)$ due to the axial symmetry of $\cosh$, hence
we can drop the $\pm$ indices in the following and only regard the positive solution $d'$, 
which is it the same time the only solution to (\ref{SE}).\\

\begin{figure}[H]
  \centering
  \includegraphics[width=0.99\textwidth, angle=0]{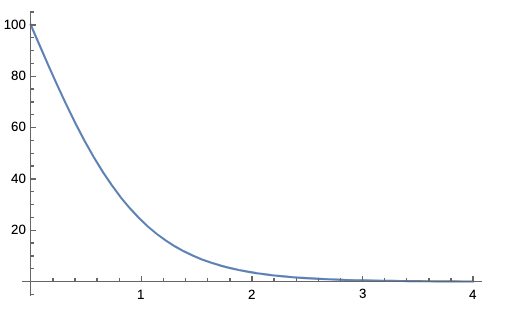}
  \caption{Realtive error in \% obtained from using (\ref{naeh}) in dependence on $d'$}
  \label{err}
\end{figure}

Equation (\ref{bcond}) has no symbolic closed form solution, but
\begin{equation} 
 \coth(d')\approx1
\label{naeh}
\end{equation} 
holds with accuracy increasing in $d'$ (see Fig.~\ref{err}).

Using (\ref{naeh}) the solution of (\ref{bcond_1}) can be approximated as 
\begin{align} 
d \approx d^0 = \frac{\kappa}{2}  = \frac{q e m R_0^2}{\hbar^2} \nonumber
\end{align}
with deviations decreasing with charge and ring size. In the Table \ref{tab1} 
some exemplary values for $R_0$ set to 1 (all atomic units) are shown. At $R_0 = 1$ bohr and $q = 1 e$  
we have $\epsilon=-d^2=-\frac{d'^2}{\pi^2}$ and $E= -\frac{d'^2 \hbar^2}{2 m \pi^2 R_0^2} \approx -0.0145$ Hartree (for comparision
in the approximation (\ref{naeh}) it yields 
$E^{0}=-\frac{{d^{0}}^2}{\pi^2} \approx -0.0127 $ Hartree (further values are listed in Table \ref{tab1}). 
In addition we note that with increasing radius $R_0$ the approximative solutions will be of
increasing quality.\\

\begin{table}[H]
\begin{tabular}{c|ccccccccccccccc}
  $q$      &  1         & 2        & 3 & 4 & 5  \\ \hline
  ${d^0}'$ &  0.5       & 1.0      & 1.5 & 2.0 & 2.5 \\
  $d'$     &  0.53575  &  1.00366  & 1.50024  & 2.00001         & 2.5 \\
  ${E^0}'$ & -0.01267  & -0.05066 & -0.11399 & -0.202642       & -0.31663     \\
  $E^0$    & -0.01454  & -0.05103 & -0.11402 & -0.202642       & -0.31663     \\
\end{tabular}
\caption{Numerically exact ($d'$) and approximated solutions (${d^0}'$) of (\ref{bcond_1}) at different total charges $q$ of the Dirac delta potential and corresponding approximate (${E^0}'$) and numerically exact energies (${E^0}'$) at $1$ Bohr radius.}
\label{tab1}
\end{table}

\begin{figure}[H]
  \centering
  \includegraphics[width=0.99\textwidth, angle=0]{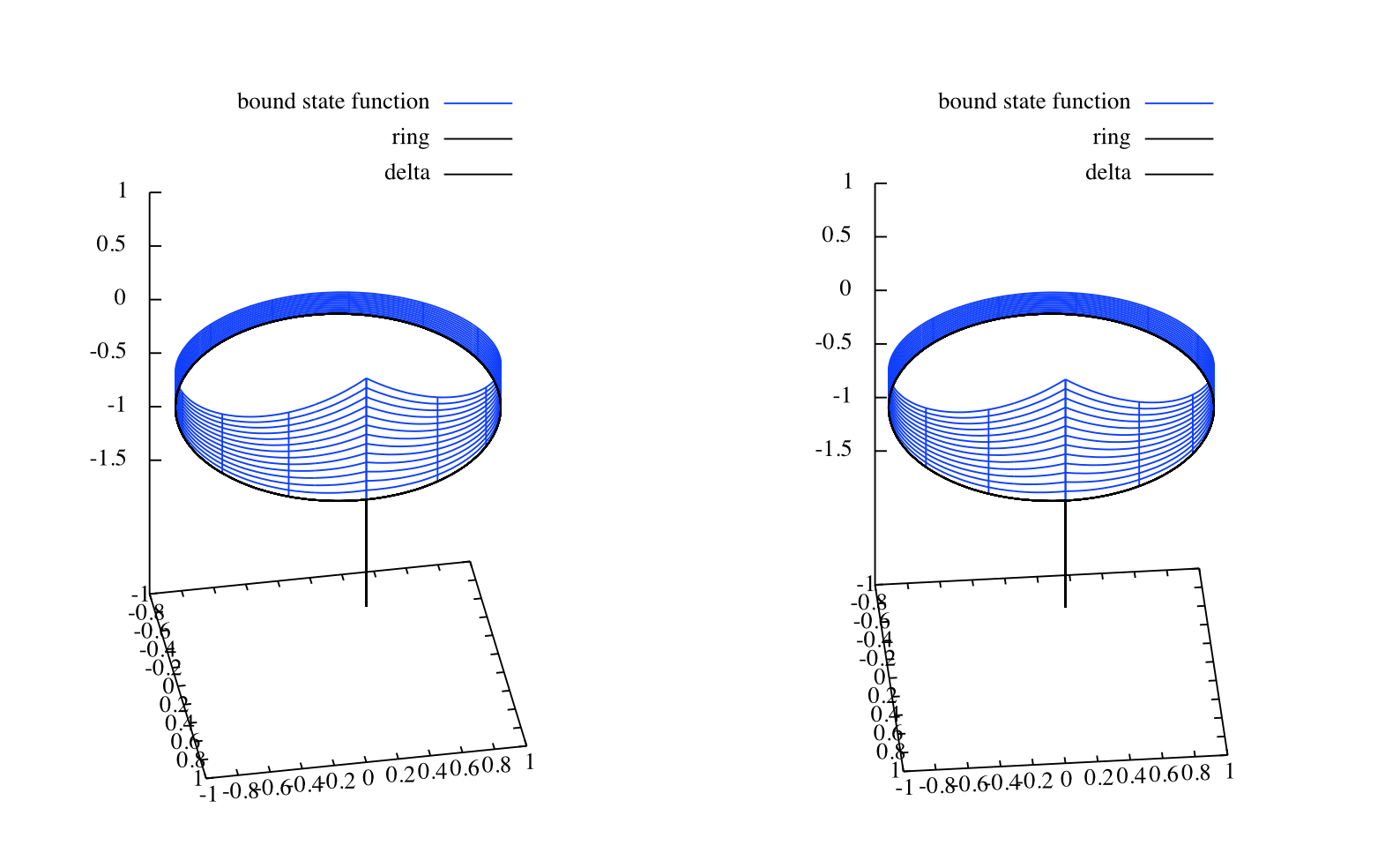}
  \caption{Stereographic projection of $\psi(\vartheta)$ for the bound state of the 
    partice in the ring with $\delta$ well.}
  \label{fig1}
\end{figure}

A graphical display of the wave function of the bound state is shown in Fig.~\ref{fig1}.

\subsection{Unbound states ($E > 0$)}

In the spirit of (\ref{ansatz}) the unbound states can be obtained from the {\em Ansatz}
\begin{align}
\psi(\vartheta;d i) = C (e^{-d i \vartheta} + e^{d i (\vartheta - 2\pi)})
\label{ansatz_unbound}
\end{align}
Yielding in analogy to (\ref{bcond_1})
\begin{align}
-2 i d + 2 i d e^{-i 2\pi d} + \kappa (1+e^{-i 2\pi d}) & = 0 \nonumber \\
\cot(\pi d) & = \frac{2}{\kappa}d
\label{bcond_2}
\end{align} 
In contrast to (\ref{bcond_1}) this yields for finite positive $\kappa$ an infinite set of solutions.
As in (\ref{bcond_1}) the system is strictly not analytically solvable in terms of a finite 
closed expression. However, for sufficiently large $d$ or large values of $\kappa$ the solutions 
are approaching 
\begin{equation}
  d_{n_{+/-}} \approx d^{(0)}_{n_{+/-}}  = \pm n \frac{\kappa}{2} \label{unbound_d}
\end{equation}

The first 5 unbound solutions for $d$ and the corresponding energies 
of the sytem with charge $+1e$ and $R=1$ bohr are given in table \ref{tab2}.

\begin{table}[H]
\begin{tabular}{c|ccccccccccccccc}
$n$             & 1       & 2       & 3       & 4       & 5       \\ \hline
$d^{(0)}_{n_{+}}$ & 0.5     & 1.0     & 2.0     & 3.0     & 4.0     \\
$d_{n_{+}}$     & 0.34278 & 1.15979 & 2.09395 & 3.06518 & 4.04963 \\
$E_n$           & 0.05875 & 0.67256 & 2.19231 & 4.69766 & 8.19976       
\end{tabular}
\caption{First five unbound solutions for $d$ and the corresponding energies 
of the sytem with charge $+1e$ and $R=1$ bohr.}
\label{tab2}
\end{table}

Since $\psi(\vartheta;d i)$ are not purely real functions, we first decompose them into 
real and imaginary part\\

\begin{align}
 \Re{[\psi(\vartheta;d i)]} & =C [\cos(d \vartheta) + \cos(d(2\pi - \vartheta))] \label{real}\\
                      & = \left( \frac{\sin (2 \pi  d)}{2 d}+\pi \right)^{-\frac{1}{2}} \cos[d(\pi - \vartheta')] \\
 \Im{[\psi(\vartheta;d i)]} & =C [\sin(d \vartheta) + \sin(d(2\pi - \vartheta))] \label{imag} 
\end{align}\\ 

and we note that (\ref{imag}) in general is dicontinuous at the $\delta$ well, thus must be rejected.
(\ref{real}) can be rewritten as a single cosine function originating at the position opposing 
the origin $\vartheta'_o = \pi$
\begin{align}
 \Re{[\psi(\vartheta;d i)]} & = 2 C \cos(d \pi)[\cos(d(\pi - \vartheta))] \nonumber \\
                      & =C'\cos[d(\pi - \vartheta')] \label{really}
\end{align}\\ 
with $-\pi \le \vartheta' < \pi$ and the normalisation constant 
\begin{align}
 C' = \left( \frac{\sin (2 \pi  d)}{2 d}+\pi \right)^{-\frac{1}{2}} \label{norm-ub}
\end{align}
In summary this yields the unbounded state functions
\begin{align}
 \psi_n & = \sqrt{ \frac{\sin (4 \pi  d_n)}{4 d_n}+\pi}^{-1} \cos[d_n(\pi - \vartheta)] \label{psiud}
\end{align}\\ 
for $-\pi \le \vartheta < \pi$ and $n>0$ with energies 
\begin{align}
E_n = -\frac{\hbar^2}{2 m R_0^2} d_n^2
\end{align}
where $d_n$ are corresponding to the positive solutions of
\begin{align}
\cot(\pi d_n) & = \frac{2\pi \hbar^2}{q e m R_0}d_n. 
\label{bcond_21}
\end{align} 
which are approximated for large $q$, $R_0$ or $n$ by 
\begin{equation}
  d_n \approx d^{0}_n  = n \frac{q R_0 e m}{2 \pi \hbar^2} \label{unbound_d_final}
\end{equation}
where we have removed the symmetry equivalent negative solutions, since $\cos$ is an even function, 
thus droped the sign indices, as compared to
(\ref{unbound_d}). Hereby we note that in comparison to the particle in the ring 
(without additional well potential) we have lost the two-fold degeneracy of the higher 
(non-ground) states. Which is an obvious consequence of the symmetry breaking due to the potential. 
%We will discuss this in more detail in the next subsection.

\subsection{$\kappa$ dependence}
The $\kappa$ dependence of the real solutions of (\ref{a}) is illustrated in the graph below.
The approximative solutions (\ref{unbound_d_final}) we have used are based on the asymptotic approximation of
the $\tan$ branches to $ x = 2n+1 $ for $n\in\mathbb{R}$.
\begin{figure}[H]
  \centering
  \includegraphics[width=0.99\textwidth, angle=0]{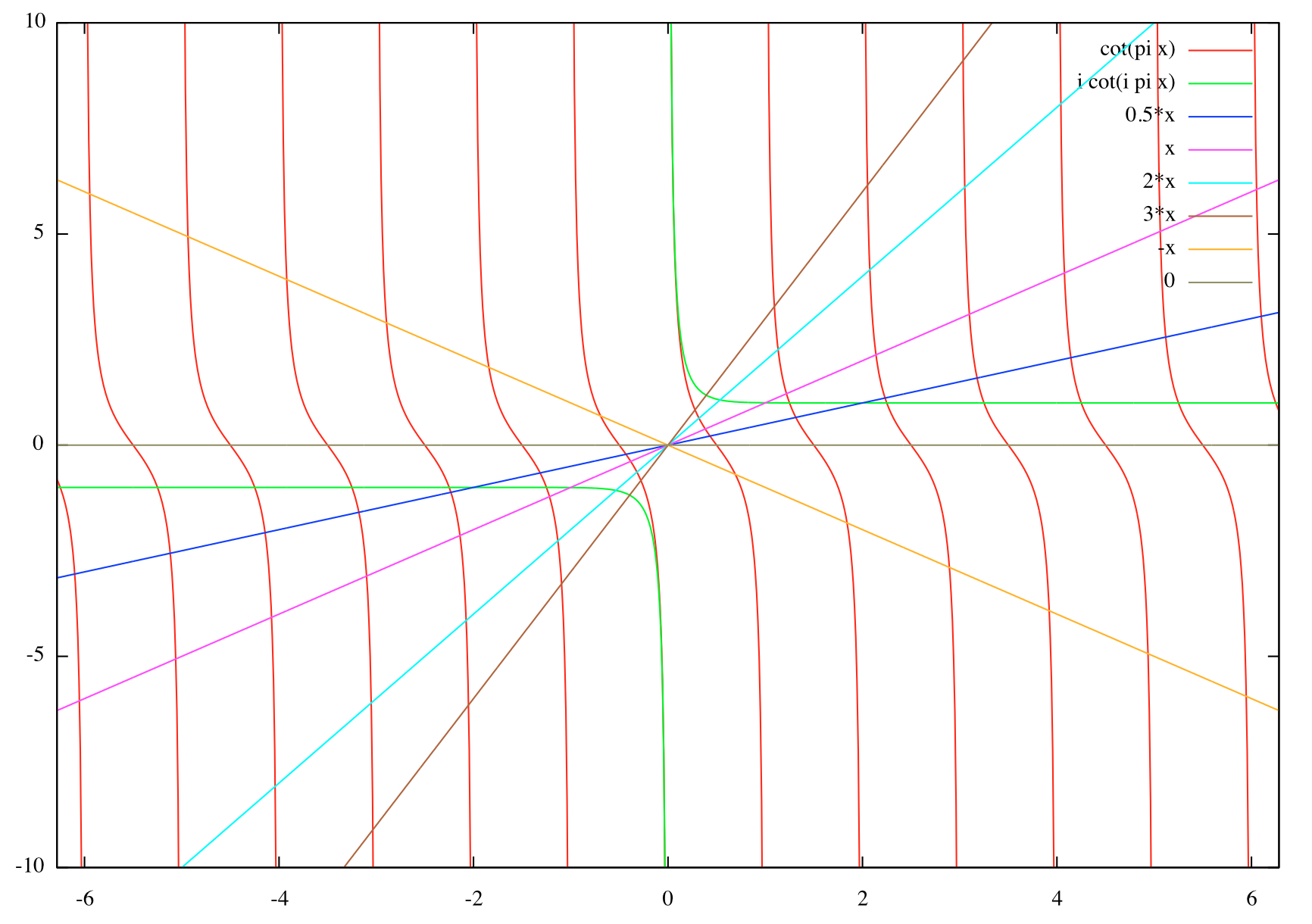}
  \caption{Graphical representation of the solutions of the characteristic equation 
(\ref{bcond_21}) for the problem with $\kappa=2$}
  \label{fig2}
\end{figure}

\section{Summary} 
Using the relations $\cos(ix)=\cosh(x)$ and $\sin(ix)=\sinh(x)$ we can combine the bound (\ref{ansatz}) 
and the unbound (\ref{psiud}) solutions to one expression

\begin{align}
 \psi_n & = \sqrt{ \frac{\sin (4 \pi  a_n)}{4 a_n}+\pi}^{-1} \cos[a_n(\pi - \vartheta)] \label{psib}
\end{align}\\

with $\kappa = q R_0 \frac{e m}{\pi \hbar^2}$, and the corresponding energies

\begin{align}
E_n = \frac{\hbar^2}{2 m R_0^2} a_n^2,\;\; \forall n\in\mathbb N_0.
\end{align}

and where $a_0$ is the (single) purely imaginary solution and $a_n$ with $n>0$ are the purely real 
solutions of the equation

\begin{equation} 
\cot(\pi a) = \frac{2}{\kappa} a \label{a}
\end{equation}

\bibliography{my} 
\bibliographystyle{rsc} %the RSC's .bst file

\end{document}